\documentclass[sn-mathphys]{sn-jnl}

\jyear{2022}%

\theoremstyle{thmstyleone}%
\theoremstyle{thmstyletwo}%

\theoremstyle{thmstylethree}%

\usepackage{graphicx}
\usepackage{float}
\usepackage{tabularx} 
\usepackage{ragged2e} 
\usepackage{booktabs} 
\newcolumntype{L}{>{\RaggedRight\hangafter=1\hangindent=0em}X}

\begin{document}

\title[Article Title]{In-orbit Performance of ME onboard \textit{Insight}-HXMT in the first 5 years}

\author[1]{\fnm{Ying} \sur{Tan}}
\author[1]{\fnm{Xuelei} \sur{Cao}}
\author*[1]{\fnm{Weichun} \sur{Jiang}}\email{jiangwc@ihep.ac.cn}
\author[1]{\fnm{Xiaobo} \sur{Li}}
\author[1]{\fnm{Bin} \sur{Meng}}
\author[1]{\fnm{Wanchang} \sur{Zhang}}
\author[1]{\fnm{Sheng} \sur{Yang}}
\author[1]{\fnm{Tao} \sur{Luo}}
\author[1]{\fnm{Yudong} \sur{Gu}}
\author[1]{\fnm{Liang} \sur{Sun}}
\author[1]{\fnm{Xiaojing} \sur{Liu}}
\author[1]{\fnm{Yuanyuan} \sur{Du}}
\author[1]{\fnm{Jiawei} \sur{Yang}}
\author[1]{\fnm{Yanjun} \sur{Xu}}
\author[1]{\fnm{Jinyuan} \sur{Liao}}
\author[1]{\fnm{Yupeng} \sur{Xu}}
\author[1]{\fnm{Fangjun} \sur{Lu}}
\author[1]{\fnm{Liming} \sur{Song}}
\author[1]{\fnm{Shuangnan} \sur{Zhang}}

\affil*[1]{\orgdiv{Key Laboratory of Particle Astrophysics}, \orgname{Institute of High Energy Physics, Chinese Academy of Science}, \orgaddress{\street{19B Yuquan Road, Shijingshan District}, \city{Beijing}, \postcode{100049},  \country{China}}}




\abstract{
\unboldmath
\textbf{Introduction:} The Medium Energy X-ray telescope (ME) is a collimated X-ray telescope onboard the \textit{Insight} hard X-ray modulation telescope (\textit{Insight}-HXMT) satellite. It has 1728 Si-PIN pixels readout using 54 low noise application-specific integrated circuits (ASICs).  ME covers the energy range of 5--30\,keV and has a total detection area of 952 cm$^2$. The typical energy resolution of ME at the beginning of the mission is 3\,keV at 17.8\,keV (Full Width at Half Maximum, FWHM) and the time resolution is 255 $\mu$s. In this study, we present the in-orbit performance of ME in its first 5 years of operation.

\textbf{Methods:} The performance of ME was monitored using onboard radioactive sources and astronomical X-ray objects. ME carries six $^{241}$Am radioactive sources for onboard calibration, which can continuously illuminate the calibration pixels. The long-term performance evolution of ME can be quantified using the properties of the accumulated spectra of the calibration pixels. In addition, observations of the Crab Nebula and the pulsar were used to check the long-term evolution of the detection efficiency as a function of energy.

\textbf{Conclusion:} After 5 years of operation, 742 cm$^2$ of the Si-PIN pixels were still working normally. The peak positions of $^{241}$Am emission lines gradually shifted to the high energy region, implying a slow increase in ME gain of $1.43\%$. A comparison of the ME spectra of the Crab Nebula and the pulsar shows that the E-C relations and the redistribution matrix 
file are still acceptable for most data analysis works, and there is no detectable variation in the detection efficiency.
}




\maketitle

\section{Introduction}\label{sec1}

The Medium Energy X-ray telescope (ME)~\cite{Cao2020ME} is one of the three main payloads onboard the \textit{Insight} Hard X-ray Modulation Telescope (\textit{Insight}-HXMT)  satellite~~\cite{Zhang2020Overview,Liu2020HE,Chen2020LE}, launched on June 15, 2017 (Modified Julian Day, MJD, 57920). ME covers an X-ray energy range of 5 to 30\,keV with a total detection area of 952 cm$^2$. The energy resolution of ME is 3\,keV (FWHM) at 17.8\,keV and the typical time resolution is 255 $\mu$s. The main features of ME are advanced among similar telescopes~\cite{hxdsuzaku,inorbitsuzaku}. Combination of ME, the High Energy X-ray Telescope (HE, 20--250\,keV, 5000 cm$^2$, timing resolution: 2 $\mu$s), and the Low Energy X-ray Telescope (LE, 1--10\,keV, 384 cm$^2$, timing resolution: 1 ms) makes \textit{Insight}-HXMT an observatory with broad X-ray energy coverage, large effective area, good energy resolution and high time resolution. Since its launch, \textit{Insight}-HXMT has obtained significant results in observational studies of X-ray binaries, magnetars, and isolated X-ray pulsars~\cite{LiCK2021,LINLIN2020,You2021,Chen_2021,Kong_2021}, in which ME plays an important role.

ME is a collimated X-ray telescope that uses Si-PIN detectors and low readout noise application specific integrated circuits (ASICs). The ME collimator consists of an aluminum alloy frame and internal tantalum inserts, that can collimate X-rays with energies up to 30\,keV. The fields of view of ME were designed to be 1$^{\circ}$ $\times$ 4$^{\circ}$, 4$^{\circ}$ $\times$ 4$^{\circ}$, 1$^{\circ}$ $\times$ 4$^{\circ}$ (full occlusion). The leakage current of the Si-PIN pixel is as low as 10 pA at the typical operating temperature of $-25$ {\textcelsius}, when the depletion voltage supplied to the electrodes. There are 1728 Si-PIN pixels, each with a pixel size of 56.25 mm$^2$ (12.5 mm$\times$4.5 mm) and thickness of 1 mm for a total detection area of 952 cm$^2$. The ASIC used in ME is an improved version of VA32TA6 (Manufacturer: GM-Idea Company)~\cite{5076101}. It is a low-power consumption integrated circuit with 32 low-noise readout channels. Thus, each ASIC can read the signals of 32 Si-PIN pixels simultaneously. Each channel consists of  low-noise charge sensitive amplifier (CSA), slow shaper, fast shaper, discriminator, sample  hold circuit, and ADC unit. Owing to the low leakage current of the Si-PIN detector and the low noise readout, the energy resolution of ME was approximately 3\,keV (FWHM) at 17.8\,keV. ME covers an energy range of 5 to 30\,keV, as determined by the thickness of the depletion region of the Si-PIN detector, noise of the system, and the dynamic range of the readout electronics. The typical dead time for one event is 255 $\mu$s, which is limited by the readout speed of the ASIC. The absolute timing accuracy is better than 10 $\mu$s.

In this study, we present the evolution of of ME performance in orbit over the past 5 years. Section \ref{sec2} introduces the current operational status of ME. The detailed in-orbit performance is presented in Section \ref{sec3}, including the energy scale, energy resolution, energy response and effects of the long-term performance evolution. Section \ref{sec4} summarizes the main results of the performance evolution study.
 
\section{Current Operation Status}\label{sec2}
To enhance the overall reliability of ME, it was designed as a modular system, with each module independent of the others. The 1728 Si-PIN pixels are installed in three medium energy detector boxes (MEDs), and each MED contains three identical detector units. All 192 pixels in a detector unit share a set of data acquisition electronics and a high voltage power supply module, which are controlled by a field programmable gate array (FPGA)~\cite{Cao2020ME,chen2010Development}. Because each pixel is read out by its own ASIC channel, it functions independently and does not affect other pixels in the event of a failure. 
 
\begin{figure}[H]
\centering
\includegraphics[width=1.0\textwidth]{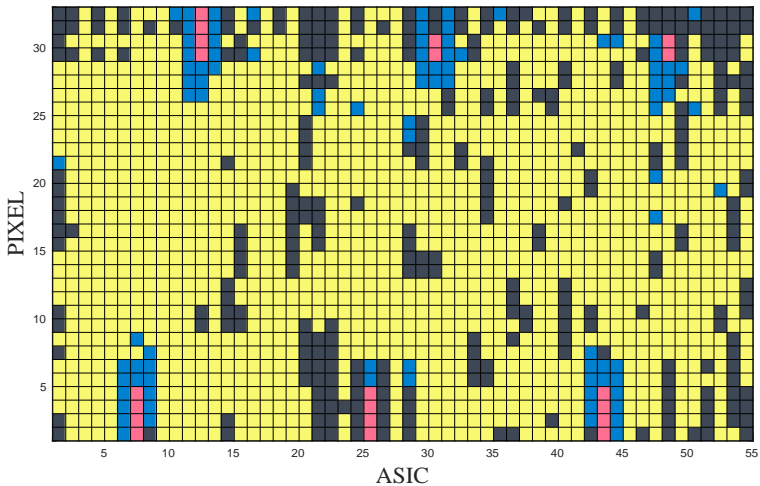}
\caption{The distribution of $1728$ pixels of ME: yellow denotes the 1319 pixels recommended for use, black denotes the 298 pixels that have been turned off, red denotes the 20 calibration pixels, and blue represents the 91 pixels that can not give reliable spectral information. The software automatically excludes the pixels of the last three categories in the standard processing flow of \textit{Insight}-HXMT.}\label{fig1}
\end{figure}

To achieve high energy resolution, the Si-PIN pixels must maintain a low leakage current during operation. Nevertheless, pollutant deposits and radiation damage degrade the detector performance by increasing the leakage current, resulting in low energy resolution and high noise count rate. In addition, radiation damage can also degrade the detector performance by decreasing the charge transfer efficiency. Every 192 pixels or channels were integrated into a single detector unit, using the same FPGA and data acquisition electronics. Consequently, a single noisy pixel affected not only the energy spectrum of the detector, but also the percentage of dead time for all 192 pixels, resulting in a decrease in the adequate photon acquisition time and detection efficiency. Owing to its modular and fault-isolating design, ME can turn off noisy pixels or channels to eliminate their influence on normal ones. Currently, 298 pixels are switched off because of excessive noise, 20 pixels are used for in-orbit calibration only, 91 pixels do not have excessive noise but cannot provide reliable spectral information for unknown reasons, and the remaining 1319 pixels operate normally and are used to observe celestial X-ray sources.  The total detection area of the 1319 pixels is 742 cm$^2$. The above statistics for the ME detector pixels are shown graphically in Fig. \ref{fig1}.  

\begin{figure}[H]
\centering
\includegraphics[width=1.0\textwidth]{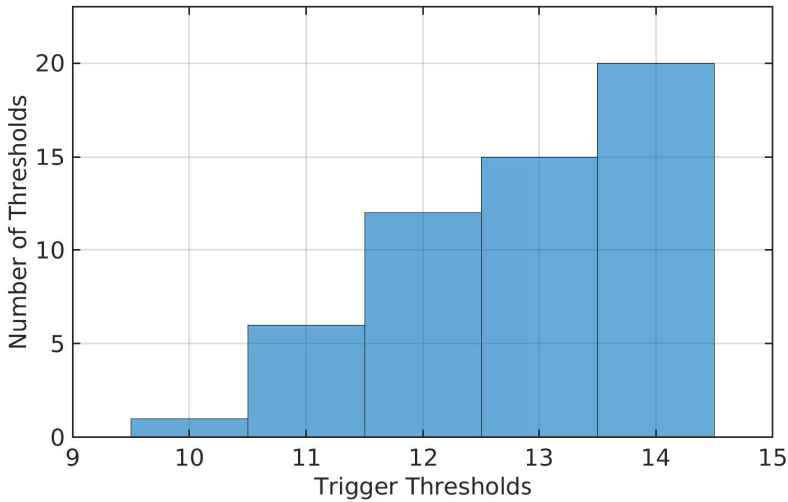}
\caption{Energy threshold distribution of 54 ASICs. }\label{fig2}
\end{figure}

The ME electronic system has two thresholds: trigger thresholds and data selection thresholds.  Fig \ref{fig2} shows the distribution of the trigger thresholds set by the 54 ASICs.  When the amplitude of the channel's fast-shaped signal with a shaping time of 300 ns~\cite{Cao2020ME} exceeded the trigger threshold, an event would be read out. Furthermore, when the ADC value of a channel's slow-shaped signal with the peaking time of 1 $\mu$s~\cite{Cao2020ME} exceeded the threshold for data selection, it would be recorded as a valid scientific event. If the ADC values of all channels did not exceed the data selection threshold during a trigger, a pseudo-event with an ADC value of 0 on channel 0 for the triggered ASIC would be recorded. Pseudo-events would be considered in the dead time counting. 

Two main parameters are involved in the trigger criteria, the trigger threshold itself  and the noise on the fast shaper. The trigger threshold is a software-programmable value that changes the common threshold for the 32 channels of the ASIC. In the on-ground calibration, we found that the measured drop in quantum efficiency close to the threshold was markedly broader than the expected sharp cutoff value. This can be explained by the noise in the fast shaper due to the short shaping time. Note that this noise term is not related to any energy resolution effect and only affects
the shape of the quantum efficiency curve at low energies. Therefore, the 5--10\,keV data were not published in scientific products, whereas the data above 10\,keV were corrected by calculating the detection efficiency curve (see Section \ref{sec34}). 

\section{ In-Orbit Performance Evolution}\label{sec3}

\subsection{Methods}\label{sec31}

The 1728 pixels' energy scales were calibrated on the ground using radioactive isotopes $^{55}$Fe, $^{57}$Co and $^{241}$Am, as well as the low X-ray calibration facility (LXCF) at the IHEP~\cite{Zhangshu_2014}. The slopes of the energy-channel (E-C) relations of all pixels increase slowly as the temperature rises from $-30$ {\textcelsius} to $-5$ {\textcelsius}, while their intercepts do not change much. The E-C relations of different pixels at different temperatures were recorded in the pre-launch CALibration Data Base (CALDB)~\cite{Li2020InflightCalibration}. 

Because there are very few celestial X-ray sources that are bright and sufficiently stable in the 10--30\,keV range, six highly collimated internal $^{241}$Am X-ray sources were used to monitor the long-term energy scale variation of ME. These radioactive X-ray sources produce X-ray lines with energies of 13.9\,keV, 17.8\,keV, 21.6\,keV, and 26.3\,keV~\cite{M0Table}, and illuminate the 24 calibration pixels continuously. Four of the calibration pixels were switched off because of excessive noise, therefore, the remaining 20 pixels were used for the in-orbit calibration. 

Every 32 pixels read out by an ASIC was installed on a copper strip with a thickness of 5 mm to form a ME detector module. A thermistor was installed in the middle of each copper strip on the side against the detector pixels, monitoring the temperature of each detector module. Temperature data were recorded in housekeeping data. The same temperature monitoring system has been used in both on-ground and in-orbit calibrations~\cite{zhangam2022}, to minimize temperature measurement errors. 

\begin{figure}[H]
\centering
\includegraphics[width=1.0\textwidth]{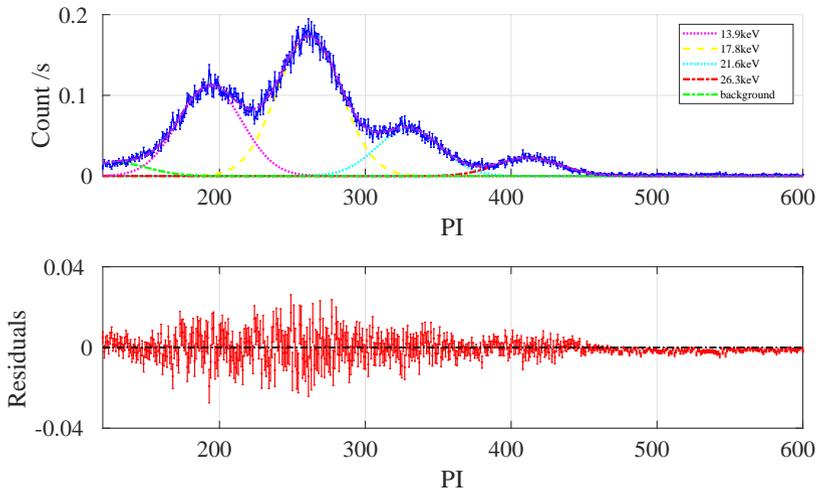}
\caption{$^{241}$\rm{Am} PI spectrum of a single calibration pixel. The dashed lines indicate Gaussian fittings for lines at 13.9\,keV, 17.8\,keV, 21.6\,keV, and 26.3\,keV, respectively.}\label{fig:am241}
\end{figure}

We can obtain instrumental pulse invariant (PI) spectra using scientific data, housekeeping data, and the pre-launch CALDB. The short-term effects of temperature variations and the differences in E-C relations between channels were corrected in the PI spectra, according to the on-ground calibration results. Therefore, the long-term performance evolution of ME can be quantified by analyzing the PI spectra of the calibration pixels at different times. By fitting the PI spectra with four Gaussian profiles, which represent the 13.9\,keV, 17.8\,keV, 21.6\,keV, and 26.3\,keV emission lines and a background continuum, the peak channels of the emission lines are obtained for individual pixels at different times to investigate the energy scale evolution, and the energy resolutions of different energies are evaluated at the same time. Figure \ref{fig:am241} shows the PI spectrum of a single calibration pixel and the fitting results. The widths of the four Gaussian functions have the same value, however, the widths can vary freely during the fitting process.

The Redistribution Matrix File (RMF) of ME was established before launch using the Geant4-based full Monte-Carlo simulation and on-ground calibration results. The energy response is not expected to change significantly in orbit even after a long time. Nevertheless, it is so important that it be accurately reconfirmed using the in-orbit observation results of the Crab nebula with the pulsar. 

We used the \textit{Insight}-HXMT Data Analysis Software package (\texttt{HXMTDAS}) and the CALDB to extract the in-orbit PI spectra of the calibration pixels and the energy spectra of the Crab nebula with the pulsar. First, we used \emph{mepical} to calculate the PI values for each event. The Good Time Interval (GTI) was then obtained using \emph{megtigen} based on the following criteria: $ ELV>10$, $COR>8$, $SAA\_FLAG=0$, $ANG\_DIST<0.04$, $T\_SAA>300$, and $TN\_SAA>300$. Second, we used \emph{megrade} to calculate the grade and the dead time of the ME events. \emph{mescreen} was used to screen the events using GTI. Finally, we obtained the spectra, response files, and background spectra using \emph{mespecgen, merspgen, and mebkgmap}, respectively.

\subsection{Energy Scale}\label{sec32}

A total of 561 pointing observations of the blank sky were selected to investigate the energy scale and energy resolution evolution, to minimize the contamination of celestial targets~\cite{ZhangJuan2020}. The total effective exposure time for these observations was 291 megaseconds (Ms). The $^{241}$Am spectra of the calibration pixels for these observations were accumulated daily. Because there were sometimes several blank sky observations on the same day, 387 spectra were finally obtained for each calibration pixel, corresponding to 387 days. The first blank sky observation was performed 110 days after launch.

\begin{figure}[H]
\centering
\includegraphics[width=1.0\textwidth]{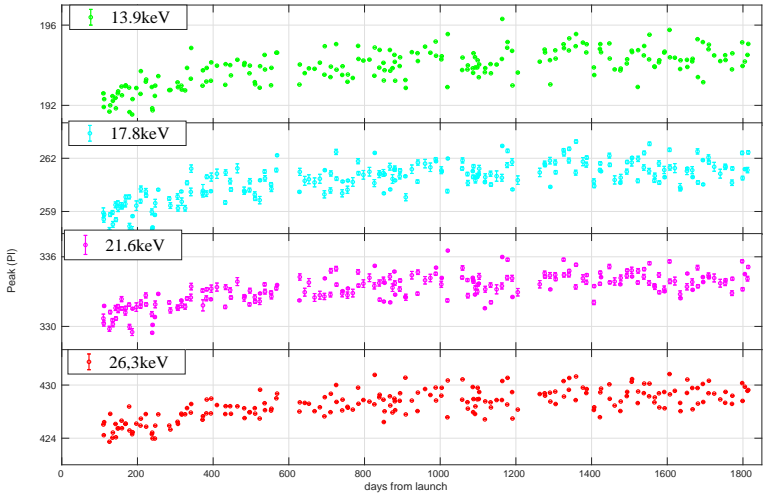}
\caption{Evolution curves of the PI peak positions of different $^{241}$\rm{Am} emission lines from a calibration pixel. The horizontal dashed lines of different colors from top to bottom correspond to 13.9\,keV, 17.8\,keV, 21.6\,keV, and 26.3\,keV, respectively. Each data point represents one day of blank sky observations accumulated.}
 \label{fig:5peak}
\end{figure}

\begin{figure}[H]
\centering
\includegraphics[width=1.0\textwidth]{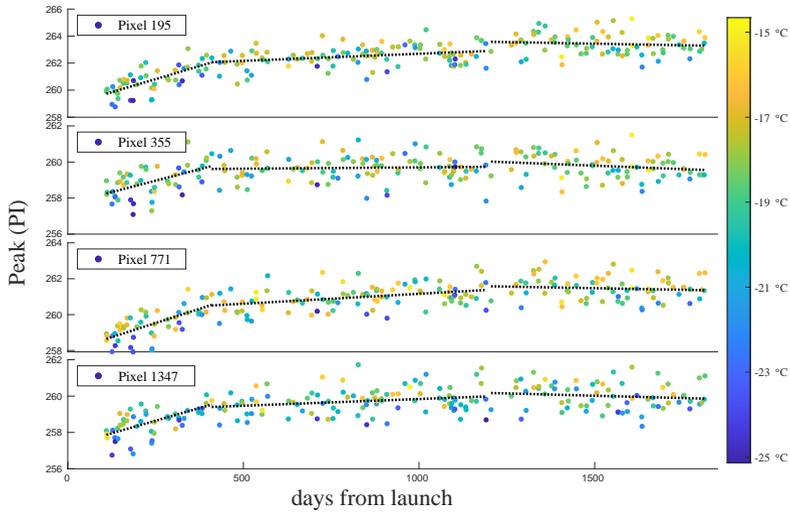}
\caption{Evolution curves of the PI peak positions of 17.8\,keV emission lines from 4 calibration pixels. Each data point represents one day. Different colors in the graph correspond to different average temperatures in the GTI. Dashed lines are the fitted linear lines for each segment.}\label{fig:slope1}
\end{figure}

\begin{figure}[H]
\centering
\includegraphics[width=1.0\textwidth]{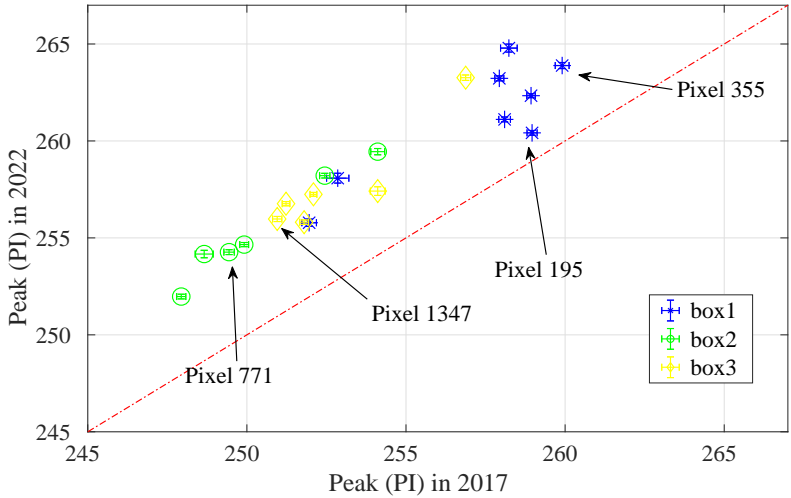}
\caption{Comparison between PI peak channels of the 17.8\,keV line for all calibration pixels in 2017 and 2022. The different colors represent calibration pixels located in different detector boxes.}\label{fig:peak2017}
\end{figure}

All 7740 PI spectra were individually fitted. We conducted a comprehensive and systematic analysis of all the fitting results and found that the peak positions of the different spectral lines of all the calibration pixels exhibited similar evolutionary trends. Therefore, we randomly selected one pixel to display the peak position variation of its four PI spectral lines in Fig. \ref{fig:5peak}, and randomly selected four pixels to display their PI peak position evolution of 17.8\,keV in Fig. \ref{fig:slope1}. Finally, Fig. \ref{fig:peak2017} presents a detailed comparison between the PI peak positions of 17.8\,keV for the 20 calibration pixels measured in 2017 and 2022. This comparison provides a general overview of the energy scale evolution of ME.

From the fitting results, we can see that the peak positions of the emission lines for all the calibration pixels increase gradually. Fig. \ref{fig:5peak} illustrates the fitted PI peak positions of the four emission lines accumulated by a calibration pixel (pixel 194, the pixel number ranged from 0 to 1727). Careful observation of the figure shows that the peak positions rose relatively rapidly from day 110 to day 410 and more slowly afterward. After day 1205, the rate of increase slowed further and even showed negative growth on some curves. A similar evolution rule can be observed in Fig. \ref{fig:slope1}, where the PI peak positions at 17.8\,keV were measured according to four randomly selected calibration pixels (Pixel 195, Pixel 355, Pixel 771, Pixel 1347). 

In Fig. \ref{fig:peak2017}, the PI peak positions of the 17.8\,keV line for the 20 calibration pixels in 2022 are plotted against those in 2017. The four pixels illustrated in Fig. \ref{fig:slope1} are indicated by the arrows in Fig. \ref{fig:peak2017}. All the peak positions showed a slight increase. The peak positions in 2022 are higher than those in 2017 by about $1.43\%$ averagely. 

There is an AC coupling in the transmission of the ME signal, therefore, an increase in the detector leakage current does not lead to an increase in the PI peak channels. In general, an increase in the gain of the readout electronics only decreases the slope of the E-C relationship without changing the intercept. To identify the cause of the change in the ME energy scale, recalculated E-C parameters were generated from the PI peak channel data and $^{241}$Am energy spectral data. We found that the slopes of the recalculated E-C relationships gradually decreased over time, but the intercepts remained stable. This suggests that the increase in the gain of the readout electronics may be responsible for this phenomenon. 

\subsection{Energy resolution}\label{sec33}

\begin{figure}[H]
\centering
\includegraphics[width=1.0\textwidth]{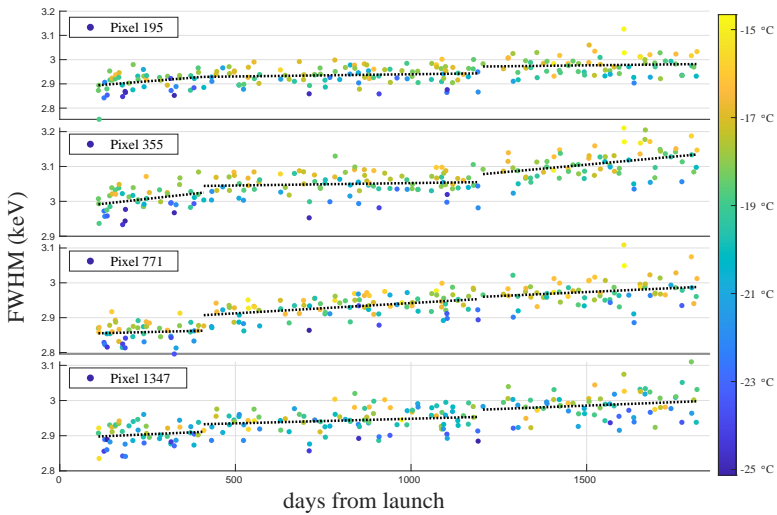}
\caption{Top: Long-term variation of the FWHMs in one calibration pixel during the past 5 years. Each data point represents one day. Different colors in the graph correspond to different average temperatures in the GTI. Dashed lines are the linear fit lines for each segment.}\label{fig:fwhm4p}
\end{figure}

\begin{figure}[H]
\centering
\includegraphics[width=1.0\textwidth]{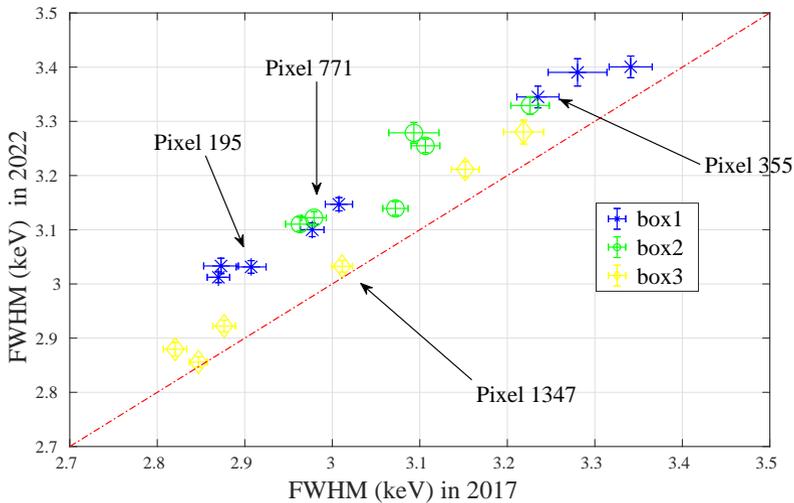}
\caption{Comparison between the energy resolution of the 17.8\,keV line for all calibration pixels in 2017 and 2022. The different colors represent the calibration pixels located in different detector boxes.}\label{fig:fwhm}
\end{figure}

When the 7740 PI spectra were fitted in Section \ref{sec32}, the FWHMs and peak channels of these Gaussian profiles could be obtained simultaneously. The FWHM in units of PI can be converted to the energy resolution in units of keV using the equation established in the on-ground calibration, $E = PI\times60/1024+3$,
where $E$ is the energy in keV, and PI is the PI channel. The energy resolution at 17.8\,keV of the calibration pixels, which is a instrumental requirement indicator for ME, was comprehensively investigated and found to have a similar evolutionary trend. Therefore, the energy resolution evolution at 17.8\,keV of the four calibration pixels (Pixel 195, Pixel 355, Pixel 771, Pixel 1347) randomly selected in Section \ref{sec32} is illustrated in Fig. \ref{fig:fwhm4p}. Finally, in Fig. \ref{fig:fwhm} we present a detailed comparison between the energy resolution of 17.8\,keV for 20 calibration pixels measured in 2017 and 2022, as a general picture of the energy resolution evolution for ME.

Fig. \ref{fig:fwhm4p} shows the gradual degradation of the spectral resolution by illustrating the FWHMs at 17.8\,keV of the four selected pixels at different working temperatures and times. To investigate the relationship between the evolution of the energy scale and energy resolution, the evolution curves of the peak position (as shown in Fig. \ref{fig:slope1}) and energy resolution (as shown in Fig. \ref{fig:fwhm4p}) of the calibration pixels were divided into three segments according to time, corresponding to days 110-410, days 410-1205 and days 1205-1814 respectively. Each segment was then linearly fitted to obtain the change rate of the gain and energy resolution of each segment. The fitting results are presented in Table \ref{tab:gainFWHM}. Although the nominal energy resolution degraded relatively rapidly on days 110-410, the actual energy resolution improved slightly during this period because the gain increased faster than the nominal energy resolution. After a period of very slow degradation, the energy resolution degraded relatively rapidly after day 1205 from launch.

\begin{table}\centering
\caption{Evolution of the variation percentage of the energy resolution, energy scale and their relationship. Peak, FWHM, and FWHM/Peak represent the gain of the energy scale, energy resolution, and the gain variation excluding the effect of gain, respectively. ppm represent parts per million. }
\begin{tabularx}{\textwidth}{ p{4.5cm} c c c c }
\toprule
  Time since launch (day)  & 110-410 & 410-1205 & 1205-1814 & 110-1814\\
\midrule
  Peak(\%) & 0.89 & 0.48 & 0.06 & 1.43 \\
  Peak per day (ppm) & 29.7 & 6.0 & 1.0 & 8.4 \\
  FWHM(\%) & 0.77 & 0.68 & 0.88 & 2.35\\
  FWHM per day (ppm) & 25.7 & 8.6 & 14.4 & 13.8 \\
  FWHM/peak(\%) & -0.12 & 0.20 & 0.82 & 0.91 \\
  FWHM/peak per day (ppm) & -4.0 & 2.5 & 13.5 & 5.3 \\
\bottomrule
\end{tabularx}\label{tab:gainFWHM}
\end{table}

Fig. \ref{fig:fwhm} plots the FWHMs at 17.8\,keV for the 20 calibration pixels in 2022 against those in 2017. The 4 pixels illustrated in Fig. \ref{fig:fwhm4p} are indicated by arrows in Fig. \ref{fig:fwhm}. Quantitatively, over the last 5 years, the typical in-orbit energy resolution of ME has degraded to $3.14$\,keV at 17.8\,keV, by an average degradation rate of $2.35\%$, including the contribution of $1.43\%$ due to the increase of readout electronic gain. Consequently, the actual degradation rate of ME’s energy resolution is about $0.91\%$.

\subsection{Energy response and detection efficiency}\label{sec34}
As a standard candle, Crab has been used by various X-ray satellites for calibration for a long time~\cite{Kirsch2005}. The photon indexes of both the nebula with pulsar~\cite{Wilson-Hodge_2011,madsen2022effective} and the pulsar~\cite{Yan_2018} are generally considered stable. The flux of the Crab pulsar exhibits a long term slow decrease~\cite{Yan_2018}, and the flux of the nebula with the pulsar exhibits a slow variation of $7\%$ across the 10--100\,keV range during the last decade~\cite{Wilson-Hodge_2011,madsen2022effective}.

The spectrum of the Crab nebula with the pulsar was modeled as a simple absorbed power law as follows:
\begin{equation}
{F(E)} = wabs \times {NE} ^{-{\Gamma } }, \label{sxgs1}
\end{equation}
where $wabs$ is the interstellar absorption using Wisconsin cross-sections, $N$ is the normalization factor, ${\Gamma}$ is the power-law photon index, and $E$ is the photon energy.

The observed PI spectrum of Crab in GTI can be modeled using the equation:
\begin{equation}
{S(PI)} ={F(E)}{\times}{ARF(E)} \times  {RMF(PI,E)} +  {BG(PI)}, \label{sxgs2}
\end{equation}
where $ARF(E)$ is the detection efficiency stored as the Ancillary Response File (ARF),  $RMF(PI, E)$ is the redistribution matrix representing the instrumental response in a given PI channel to a photon with energy $E$, and $BG(E)$ is the background rate in the GTI~\cite{GUO202044}.

\begin{figure}[H]
\centering
\includegraphics[width=0.6\textwidth,angle=270]{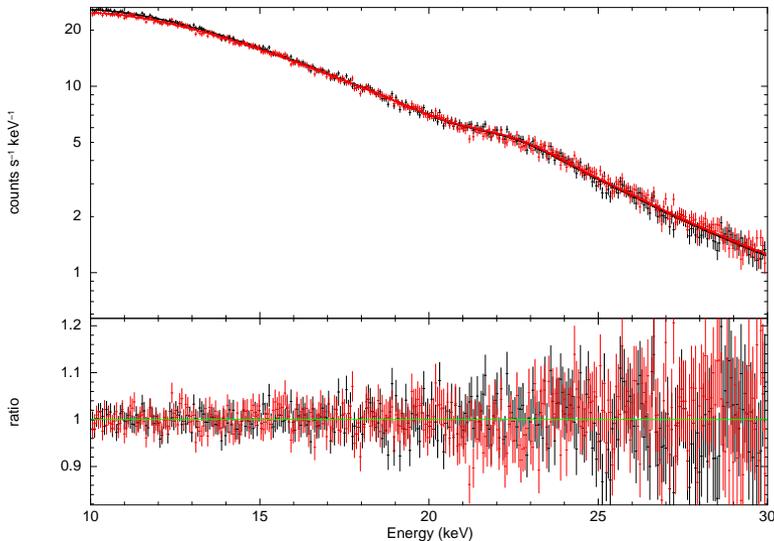}
\caption{Top: ME spectra of the Crab nebula with the pulsar for the first observation in 2017 (black) and the last in 2022 (red), with wabs $\times$  power law models. Bottom: Ratios of data (points in the top panel) to model (lines in the top panel).}\label{figcrab}
\end{figure}

To date, the RMF has remained unchanged from its original pre-launch release. Early in orbit, the ARF was generated using on-ground calibration results and Monte-Carlo simulations. The current version of the ARF was precisely calibrated and fixed by utilizing the observation results of Crab nebula with the pulsar from September 2017 to April 2019, with the calibration standard of ${\Gamma } = 2.11 $ and ${{N} } = 8.76\ {\rm  keV^{-1}cm^{-2}s^{-1}} $~\cite{Li2020InflightCalibration}, and released in December 2019.  

To simplify the background model, only the data of the narrow field of view pixels recommended in the user's manual~\cite{2018SPIElixb} were selected for the analysis of the Crab observation results. To reconfirm the RMF and ARF, we compared the first and last Crab observations of ME. Fig. \ref{figcrab} illustrates the spectra of the Crab nebula with the pulsar for the first observation performed in 2017 (black) and the last one performed in 2022 (red) fitted with the absorbed power law model in \texttt{Xspec}, 
where the model is constructed with the pre-launch RMF and the in-orbit calibrated ARF, and the column density for the interstellar absorption is fixed at $0.36 \times10^{22}\ \rm cm^{-2}$~\cite{Weisskopf_2000,refId0}. 

The fitting results for the first observation are ${\Gamma } = 2.12\pm 0.06 $ and ${{N} } = 8.87\pm 0.14 $, and the results for the last one are ${\Gamma } = 2.06\pm 0.06 $ and ${{N} } = 7.95\pm 0.16 $. There is a deviation of $2.37\%$ in $\Gamma$ from the result of the last observation to the calibration standard. Although the deviation between the median value of $\Gamma$ and 2.11 is within the error range, it is found that $N$ and $\Gamma$ are strongly correlated in the fitting process (as shown in Fig. \ref{figcrabindexvsnorm}), and a slight change of $\Gamma$ will lead to a sharp change in the value of $N$. Therefore, the $9.25\%$ deviation of $N$ may mainly come from the random deviation of the fitting result of $\Gamma$. We will conduct a systematic analysis on $\Gamma$ and $N$ to confirm the validity of ARF and RMF in the following section.

\subsection{Long-term effects and systematic errors}\label{sec35}

\begin{figure}[H]
\centering
    \includegraphics[width=1.0\textwidth]{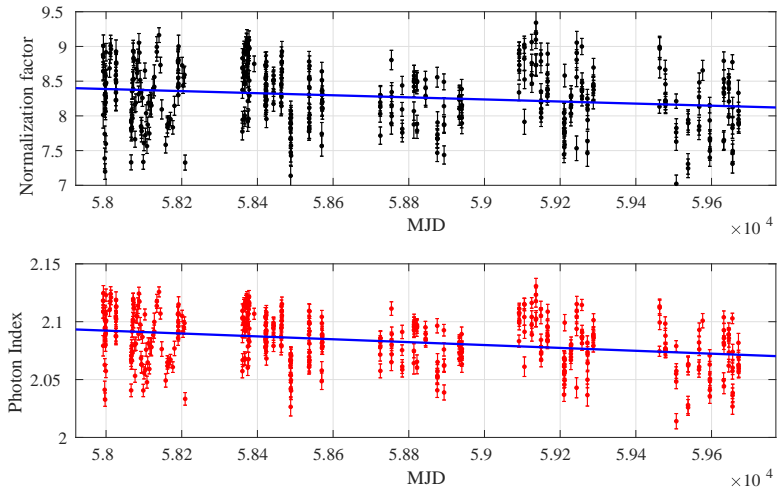}
\caption{Long-term variation of the normalization factor (top) and photon index (bottom) obtained by fitting the spectra of the Crab nebula with the pulsar in different exposures of the past 5 years. }\label{figcrabindex}
\end{figure}

\begin{figure}[H]
\centering
    \includegraphics[width=1.0\textwidth]{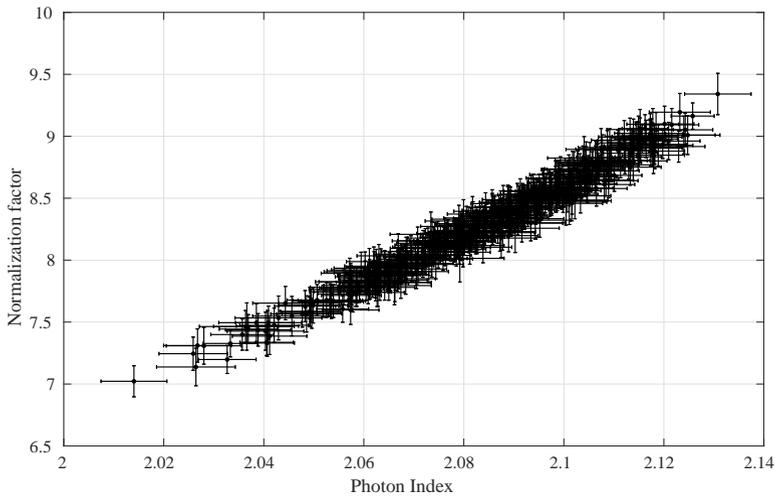}
\caption{The correlation between the normalization factor and photon index in the fitting results. }\label{figcrabindexvsnorm}
\end{figure}

In order to investigate the long-term evolution of photon index $\Gamma$ and normalized factor $N$, all 454 observations of the Crab nebula with the pulsar are analyzed, corresponding to a total effective exposure time of 1.53 Ms.

The long-term variations in the normalization factor (top) and photon index (bottom) are illustrated and linearly fitted in Fig. \ref{figcrabindex}. There are obvious random variations for $\Gamma$ and $N$, even between different observations on the same day. This may be due to the sensitivity of ME, as a collimated telescope, to the space environment, and other unknown reasons that need to be determined through a deeper understanding of the instruments and observations. The fitting results show that $\Gamma$ and $N$ decreased by $1.07\pm0.01\%$ and $3.37\pm0.02\%$ over the last 5 years, respectively. 

Over the last more than 10 years the Crab has been observed, the photon index has remained stable at ${\Gamma } \sim 2.1$~\cite{Wilson-Hodge_2011,madsen2022effective}. Therefore, we attribute that the decrease of  $\Gamma$ to the long-term co-evolution of ME gain (E-C) and energy resolution (RMF), as described in detail in Sec. \ref{sec32} and Sec. \ref{sec33}. For example, the increase in gain would lead to a decrease in $\Gamma$ by stretching the x-axis of the energy spectrum, and the degradation of energy resolution would lead to the same result by flattening the spectra. For the same reason, a similar decrease in the photon index of the Crab pulsar spectrum was observed with ME, which was not detected with LE or HE~\cite{zhao2022}. 

\begin{figure}[H]
\centering
    \includegraphics[width=1.0\textwidth]{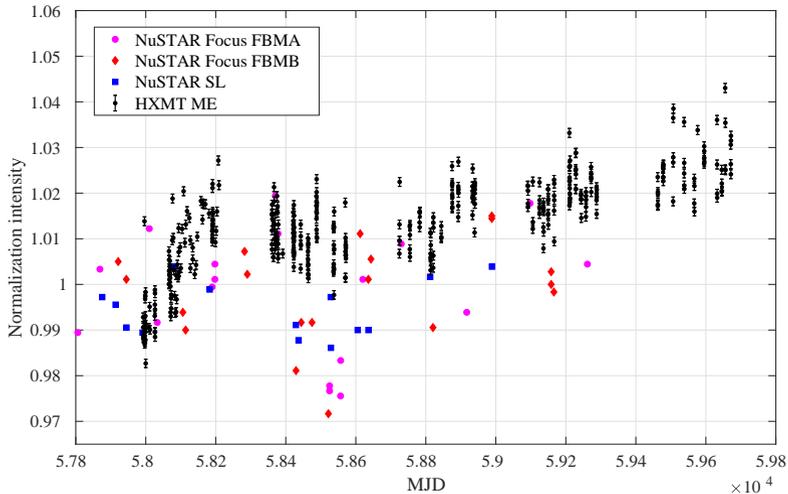}
\caption{The data points with error bars in black is the flux lightcurve of the 10--30\,keV from 454 Crab exposure observations of the ME over the last 5 years. These fluxes are normalized to the wabs $\times$  powerlaw model used for the 2019 in-orbit calibration with $N_{\rm H}=0.36 \times10^{22}\ \rm cm^{-2}$, $\Gamma=2.11$, and $N=8.76$. For comparison, the red, pink, and blue data points are the \textit{NuSTAR} 3-10 kev fluxes over the past five years, normalized to  the 3--10\,keV flux of the canonical Crab model as $1.57\times10^{-8}\ {\rm ergs\ cm^{-2}\ s^{-1}}$. }
\label{figcrabnorm}
\end{figure}

Therefore, we believe that the long-term decrease in the normalization factor $N$ is not secular or intrinsic to either the source or is caused by a change in the efficiency of the detector. This can be further understood by the strong correlation between $\Gamma$ and $N$ shown in Fig. \ref{figcrabindexvsnorm}, which shows that the normalization factor is heavily influenced by $\Gamma$ during the fitting process. Therefore, we refit the data for the 454 exposures with the wabs $\times$  powerlaw model, with the spectral index $\Gamma$ fixed at 2.11, and obtain the normalization factor over time. To discern the possible contributions to the variation of the normalization factor over time at $\Gamma= 2.11$ from the evolution of the ARF and the intrinsic variation of the Crab nebula with the pulsar, we compare the observations of ME and \textit{NuSTAR} in the same period. This is because, in the wabs $\times$ powerlaw model, the normalization factor and the flux are linearly correlated. We compared the fluxes in the 10--30\,keV band for ME in Fig. \ref{figcrabnorm} with those in the 3--10\,keV band for the \textit{NuSTAR} focus and stray-light (SL) observations ~\cite{madsen2022effective} during the same period. The fluxes of ME are normalized according to the model flux of wabs $\times$  powerlaw at $N=8.76$  and  $\Gamma =2.11$. The \textit{NuSTAR} data are normalized to the canonical flux of $1.57\times10^{-8}\ {\rm ergs\ cm^{-2}\ s^{-1}}$ between 3--10\,keV. The long term \textit{NuSTAR} flux curve shows a slow variation on the order of $1-2\%$ per year, which is on the same order of magnitude as ME; however, the \textit{NuSTAR} results on Crab have not been reported towards late times when ME has detected an apparent flux increase. 

There was a significant random fluctuation in ME fluxes on the short-term scale. By comparing the results of \textit{NuSTAR} for the same period, we found that there was no correlation between the fluxes measured by ME and \textit{NuSTAR}. Therefore, the possibility of fluctuations from Crab itself was excluded. We believe that such short-term scale fluctuations may originate from the systematic error of ME, which will be further analyzed in our next paper. Therefore, the detection efficiency of ME remains stable, and the ARF still dose not need to be recalibrated.

Consequently, the long-term evolution of ME has resulted in the current E-C relations and RMF being unable to describe instrumental features as accurately as before. As discussed above, the changes of E-C relations and RMF only resulted in about 1\% and 3\% long term variations of the photon index $\Gamma$ and the normalization factor $N$; the latter is smaller than the approximately $10\%$ absolute efficiency uncertainties that X-ray instruments typically have. Therefore, we consider the current E-C relations and RMF to be acceptable for most data analyses work. Meanwhile, re-calibration of the E-C relations and RMF for ME is planned to obtain more accurate observational results.

\section{Conclusion}\label{sec4}
Since its launch, ME onboard \textit{Insight}-HXMT has performed well and has been quite stable. The current detection area of the Si-PIN pixels was 742 cm$^2$. The measured peak positions of the $^{241}$Am emission lines show that the energy scale gain of ME increased by $1.43\pm0.01\%$, while the degradation in energy resolution was about $0.91\%$ after excluding the effect of gain. Although slight long-term effects have been found with Crab observations, both the pre-launch E-C relations and RMF are still reliable and acceptable for most data analyses work. The detection efficiency remained stable and there was no detectable deviation in the ARF.

\section*{Declarations}
\begin{itemize}
\item This study used data from the \textit{Insight}-HXMT mission, a project funded by the China National Space Administration (CNSA) and the Chinese Academy of Sciences (CAS). We gratefully acknowledge the support from the National Program on Key Research and Development Project (Grant No.2021YFA0718500) from the Ministry of Science and Technology of China (MOST). The authors thank supports from the National Natural Science Foundation of China under Grants 12273043, U1838201, U1838202, U1938109, U1938102, U1938108, and U2038109. This work was partially supported by the International Partnership Program of the Chinese Academy of Sciences (Grant No.113111KYSB20190020))
\item \textbf{Conflict of interest}  On behalf of all authors, the corresponding author states
that there is no conflict of interest.
\end{itemize}

\bibliography{sn-bibliography}

\end{document}